\documentclass[
 aip,
 reprint,]{revtex4-1}
\usepackage{graphicx}
\usepackage{dcolumn}
\usepackage{bm}
\usepackage[utf8]{inputenc}
\usepackage[T1]{fontenc}
\usepackage{mathptmx}
\usepackage{etoolbox}
\usepackage{color}
\usepackage[normalem]{ulem}
\usepackage{mathrsfs}
\usepackage{placeins} 
\usepackage{graphics}
\usepackage{bm}
\usepackage{subfigure}
\usepackage{graphicx}
\usepackage{amsthm}
\usepackage{notes2bib}
\usepackage{amsmath}
\usepackage{amssymb}
\usepackage{url}
\usepackage{enumerate}
\usepackage{color}
\usepackage{mathtools}
\usepackage{setspace}

\usepackage{color}

\newcommand{\allblack}{\color{black}
{}}

\newcommand{\mb}{\mathbf}

\usepackage{comment}
\makeatletter
\def\@email#1#2{%
 \endgroup
 \patchcmd{\titleblock@produce}
  {\frontmatter@RRAPformat}
  {\frontmatter@RRAPformat{\produce@RRAP{*#1\href{mailto:#2}{#2}}}\frontmatter@RRAPformat}
  {}{}
}%
\makeatother
\begin{document}

\preprint{AIP/123-QED}

\title[On the attractor in a high-dimensional neural network dynamics of reservoir computing: Lyapunov analysis viewpoint]{On the attractor in a high-dimensional neural network dynamics of reservoir computing: Lyapunov analysis viewpoint}
\author{Miki U. Kobayashi}
\affiliation{Faculty of Economics, Rissho University, 4-2-16 Osaki, Shinagawa-ku, Tokyo 141-8602, Japan}
\author{Kengo Nakai}
\affiliation{Graduate School of Environment, Life, Natural Science and Technology, Okayama University, Okayama 700-8530, Japan}
\author{Yoshitaka Saiki} \author{Natsuki Tsutsumi}
\affiliation{Graduate School of Business Administration, Hitotsubashi University, Tokyo 186-8601, Japan\\
(Electronic mail: miki@ris.ac.jp)}

\date{\today}

\begin{abstract}

Recent theoretical developments of reservoir computing have clarified a sufficient condition about which reservoir computing can capture the dynamics of a target system, enabling the reconstruction of dynamical invariants. %While previous studies have suggested that a reservoir capable of accurately reproducing target dynamics should exhibit matching its Lyapunov spectrum, complete numerical recovery of all Lyapunov exponents, including stable negative ones, has yet to be demonstrated.
Even when the condition is relaxed,
the reservoir computing is found to succeed in reconstructing time series.
In this study, we investigate numerically the dynamical structures underlying the embedding structure by comparing the Lyapunov spectrum of a high-dimensional neural network in a reservoir computing model with that of the actual system. We also compute Lyapunov exponents restricted to the tangent space of the inertial manifold in a high-dimensional neural network. Our results provide numerical evidence that reservoir computing can accurately identify the Lyapunov spectrum of the target system, including all negative exponents.
\end{abstract}

\maketitle

\begin{quotation}
Reservoir computing has emerged as a powerful and efficient machine learning framework for modeling and predicting complex dynamical systems. A key factor underlying its success is the ability of the reservoir’s internal states to form a stable functional representation of the target system’s dynamics. This effective representation enables the reservoir to learn and reproduce intricate behaviors directly from data. However, the dynamical structures that enable such representations, particularly for general dynamical systems, remain insufficiently understood.
In this study, we compare the stability properties—quantified by the Lyapunov exponents—of reservoir networks with those of the actual system. Our numerical results under practical settings clarify the mechanism by which reservoir computing captures the full dynamical structure of an actual system, thereby strengthening the theoretical foundation for its predictive capabilities.
\end{quotation}

\section{Introduction.}
Recently, the interest in modeling chaotic dynamics using machine learning techniques has increased~\cite{nakajima2021-book,tsutsumi22,tsutsumi23}.
Reservoir computing, a neural network-based approach, has garnered considerable attention for its efficiency~\cite{tanaka_2019, panahi_2024, kong_2023, plat_2021}.
Reservoir computing only learns the output linear map from data, which requires a low computational cost.
A data-driven model using reservoir computing 
predicts short-term trajectories and reconstructs invariant sets, such as attractors~\cite{Lu_2018}.
Even with short training trajectories, these models can accurately reproduce invariant sets~\cite{kobayashi_2021}.
Beyond invariant sets, such as fixed points and periodic orbits, 
reservoir computing reconstructs Lyapunov exponents and manifold structures, including stable and unstable manifolds, providing richer dynamic information than the training data alone.
A single data-driven model can infer time series of a macroscopic variable of chaotic fluid flow from various initial conditions and recover accurate long-term statistical properties from a single simulated trajectory~\cite{nakai_2020}.

Theoretical studies have shown that a reservoir computing model is described as an embedding on a low-dimensional manifold~\cite{hart_2024}.
Furthermore, conditions for generalized synchronization between source dynamics and reservoir dynamics in continuous-time reservoir computing have been identified~\cite{Zhixin_2020}.
To elucidate the underlying mechanism, recent studies have explored the learning process of reservoir computing~\cite{berry_2023, berry_2025}.
Recently, linear reservoir models are proved to preserve the metric structure  including angles and lengths, as well as the topological structure~\cite{Hart_2025}. 
In addition, statistics for differential topological properties
between datasets are proposed for obtaining numerical evidence of the embedding, and they are applied to polynomial reservoir computer~\cite{pecora_2025}.   
These insights form a theoretical basis for understanding reservoir computing's ability to reconstruct and predict complex dynamics.

To clarify the dynamical system structure of embedding theory and generalized synchronization in reservoir computing with practical settings, we can compare Lyapunov exponents in the reservoir space and those in the actual space, as detailed in Section~\ref{subsec:General_setting}.
It is essential to analyze dynamics on an inertial manifold and to compare the Lyapunov exponents with those of the actual models by controlling the transversal stability.

Reservoir computing models can reconstruct the non-negative Lyapunov exponents of the original dynamics in the reservoir space ~\cite{Pathak_2017,kobayashi24}.
Some studies~\cite{Pathak_2017, Pantelis_2020} also report reconstruction of negative Lyapunov exponents.
Lyapunov exponents of the constructed model in the actual space, including negative ones, align with those of the original dynamics~\cite{kobayashi_2021}.
The spectral radius, a key hyperparameter related to the echo state property, is critical for reconstructing negative Lyapunov exponents.
When the spectral radius $\rho$ is small, the negative Lyapunov exponents are reconstructed in the reservoir space~\cite{hart24}.

In this study, we investigate the relationship between the spectral radius $\rho$ and Lyapunov exponents in the reservoir space, numerically validating the embedding structure of constructed models using (covariant) Lyapunov vector\cite{ginelli_2007}, particularly for models with a non-small spectral radius.
By leveraging this geometric structure, we reproduce the original dynamics' Lyapunov exponents in the reservoir space by restricting the calculation to the low-dimensional manifold.

\section{Reservoir computing}
The reservoir is trained by inputting a time series $\mb{u}(t)$ and fitting
a linear function $\mb{W}_{\text{out}}$ of the reservoir state vector $\mb{r}(t)$, such that $\mb{W}_{\text{out}}\mb{r}(t)\approx\mb{u}(t)$~\cite{Jaeger_2001,Jaeger_2004}.

\subsection{General setting}
\label{subsec:General_setting}
Let $L_{0}$ denote the transient time and $L$ the training time.  
We assume an observable vector-valued variable $\mb{u}(t) \in \mathbb{R}^{M}~(-L_{0}\le t \le L)$.
For an observed time-series data $\{\mb{u}(t)\}$, the reservoir state vector $\mb{r}(t) \in \mathbb{R}^{N}~(N \gg M)$ 
is determined by 
\begin{equation*}
        \mb{r}(t+1)=(1-\alpha)\mb{r}(t)+\alpha \tanh(\mb{A}\mb{r}(t)+\mb{W}_{\text{in}}\mb{u}(t) + \xi{\bf 1}),
\end{equation*}
where $\mb{A} \in \mathbb{R}^{N\times N}$
and $\mb{W}_{\text{in}}\in \mathbb{R}^{N\times M}$ are matrices;
$\alpha$ ($0<\alpha\le 1$) is a coefficient;
$\xi$ is a bias parameter;
${\mb 1}= (1,1,\ldots,1)^{\text{T}}\in \mathbb{R}^{N}$.
We define $\tanh(\mb{q})=(\tanh(q_{1}), \tanh(q_{2}),\ldots,\tanh(q_{N}))^{\text{T}}
,$ for a vector $\mb{q}= (q_{1},q_{2},\ldots,q_{N})^{\text{T}}$, where
$\text{T}$ represents the transpose of a vector.
Note that $\mb{A}$ is a random matrix which has a spectral radius $\rho$; 
$\mb{W}_{\text{in}}$ is also a random matrix, each row of which has one non-zero element, chosen from a uniform distribution on $[-\sigma, \sigma ]$.

For the given random matrices $\mb{A}$ and $\mb{W}_{\text{in}}$,
we determine $\mb{W}_{\text{out}}$, such that the following quadratic form
takes the minimum: 
\begin{equation*}
    \displaystyle\sum^{L}_{l=0}\|\mb{W}_{\text{out}}\mb{r}(l)-\mb{u}(l)\|^{2}
    +\beta[Tr(\mb{W}_{\text{out}}\mb{W}^{\text{T}}_{\text{out}})],\label{eq:minimize}
\end{equation*}
where $\|\mb{q}\|^{2}=\mb{q}^{\text{T}}\mb{q}$ for a vector $\mb{q}$.
The minimizer is
\begin{align*}
    \mb{W}^{*}_{\text{out}} & =\delta\mb{U}\delta\mb{R}^{\text{T}}(\delta\mb{R}\delta\mb{R}^{\text{T}}+\beta\mb{I})^{-1}, \label{eq:wout-c1}
\end{align*}
where $\mb{I}$ is the $N \times N$ identity matrix, $\delta\mb{R}$ (respectively,
$\delta\mb{U}$) is the matrix whose $l$-th column is $\mb{r}(l)$ (respectively,
$\mb{u}(l)$). 
(see Lukosevicius and Jaeger~\cite{Lukosevicius_2009},~P.140 and Tikhonov and Arsenin~\cite{Tikhonov_1977}, Chapter 1 for details).
Using the matrix $\mb{W}^{*}_{\text{out}}$, we obtaine the reservoir computing model:
\begin{equation}
        \mb{r}(t+1)=(1-\alpha)\mb{r}(t)+\alpha \tanh(\mb{A}\mb{r}(t)+\mb{W}_{\text{in}}\mb{W}^{*}_{\text{out}}\mb{r}(t) + \xi{\bf 1}).\label{eq:reservoir}
\end{equation}
For $t>L$, predicted data $\mb{u}(t)$ is replaced by $\mb{W}^{*}_{\text{out}}\mb{r}(t)$.

The $M$-dimensional space to which the vector $\mb{u}(t)$ belongs is called the actual space, and the $N$-dimensional space to which the vector $\mb{r}(t)$ belongs is called the reservoir space. 

\subsection{Setting for our paper}
In this study, we set $\mb{A}:= \rho \mb{A^\prime}$, where $\mb{A^\prime}$ is a fixed random matrix with a spectral radius of one, except for Fig.~\ref{fig:2nd-lyap-sub}. 
Thus, the spectral radius of $\mb{A}$ is $\rho$. 
For each $\rho \mb{A^\prime}$ and a fixed random matrix $\mb{W}_{\text{in}}$, we determine the matrix $\mb{W}^{*}_{\text{out}}$. 
We focus on the reproducibility of the dynamical system structure with respect to the spectral radius.
Parameter values used in the modeling are listed in Table I.
%~\ref{tab:parameter}.
Additional details on reservoir computing can be found elsewhere~\cite{Pathak_2017,nakai_2018}.

\begin{table}[tb]
    \allblack \footnotesize
    \begin{center}
        \begin{tabular}{|l|l|r|r|}
            \hline
            % \multicolumn{2}{|c|}{model} & H\'enon                                              \\
            % \hline
            ~$M$                        & dimension of input and output variables             & 2               \\
            \hline
            ~$N$                        & dimension of reservoir state vector                 & 40              \\
            \hline 
            ~$L_0$&number of iterations for the transient& 1000  \\ \hline
            ~$L$                        & number of iterations for the training               & $2000000$        \\
            \hline
            ~$\rho$                     & spectral radius $\rho$ of $\mb{A}$                      & $0.001\sim 0.2$ \\
            \hline
            ~$\sigma$                   & scale of input weights in $\mb{W}_{\text{in}}$      & 1               \\
            \hline 
            ~$\alpha$                   & nonlinearity degree in a model~(\ref{eq:reservoir}) & 1               \\
            \hline
            ~$\xi$                      & bias parameter in a model~(\ref{eq:reservoir})      & 1             \\
            \hline
            ~$\beta$                    & regularization parameter                            & $10^{-8}$       \\
            \hline
            %~$\Delta \tau$              & delay-time for input and output variables           & 1               \\
            %\hline
        \end{tabular}
        \caption{
        {\bf The list of parameters and their values used in the reservoir computing.}
        We set the spectral radius $\rho$ for each model.
        }
    \end{center}   \label{tab:parameter}
\end{table}

\section{Reservoir computing for the H\'enon dynamics}
We deal with the H\'enon map with a set of classical parameter values:
\[
    \begin{cases}
        x_{n+1}= 1 - 1.4\, x_{n}^{2}+ y_{n} \\
        y_{n+1}= 0.3\, x_{n}.
    \end{cases}
\]
The map will be denoted as
the actual H\'enon map. 
Data-driven models were constructed using $\{\mb{u}(t)\}$, which represents the time-series data $\{(x_{n}, y_{n})^T\}$ obtained from the H\'enon map.

\begin{figure}
    \centering
    \hspace{-1mm}
    \includegraphics[width=0.35\columnwidth, height=0.4\columnwidth]{
        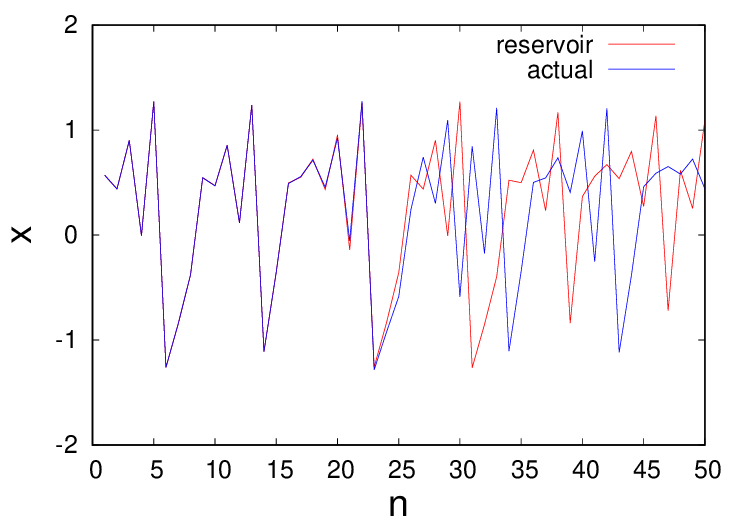
    }
    \includegraphics[width=0.35\columnwidth, height=0.4\columnwidth]{
        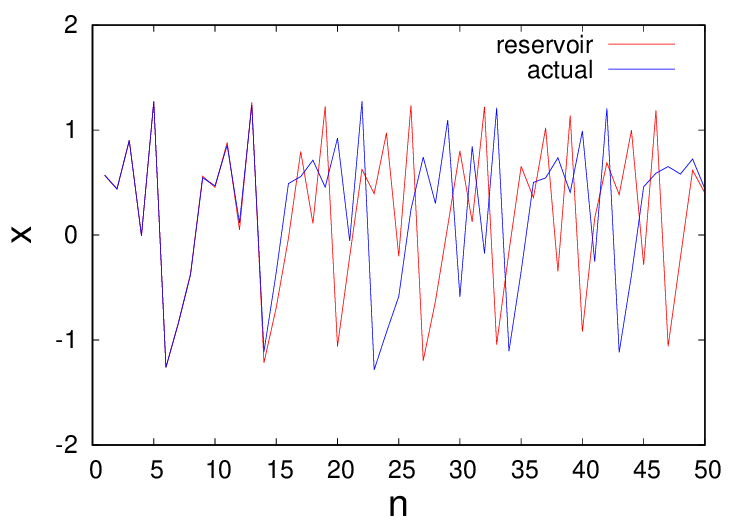
    }
    \includegraphics[width=0.50\columnwidth]{
        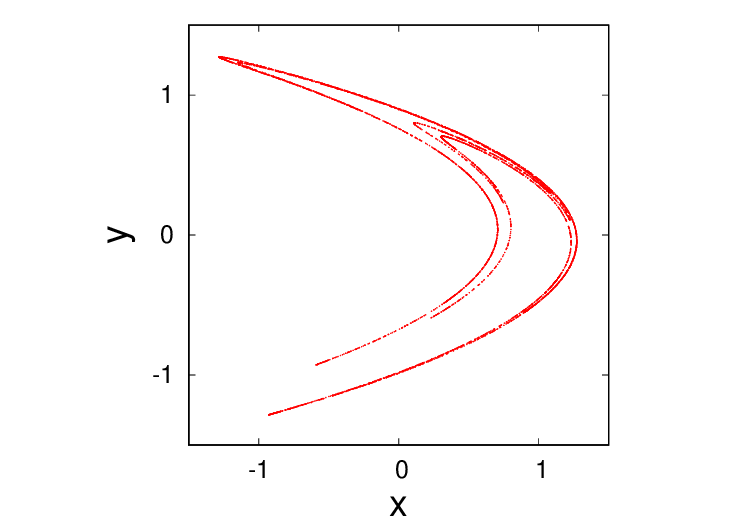
    }\hspace{-13mm}
    \includegraphics[width=0.50\columnwidth]{
        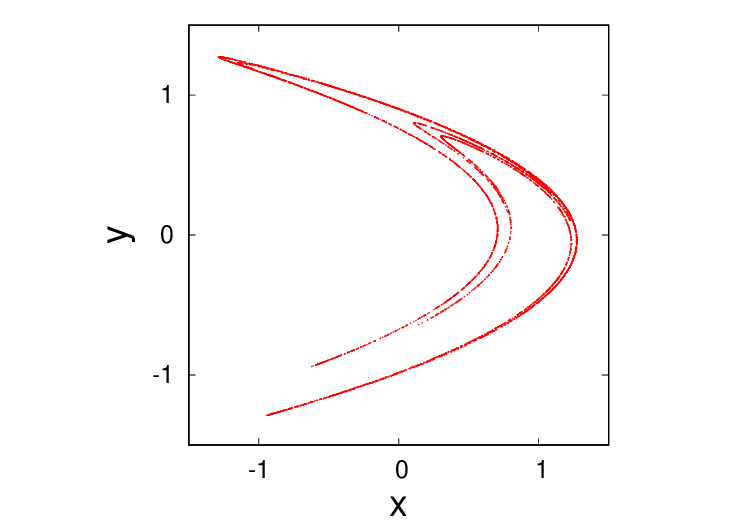
    }
    \includegraphics[width=0.50\columnwidth]{
        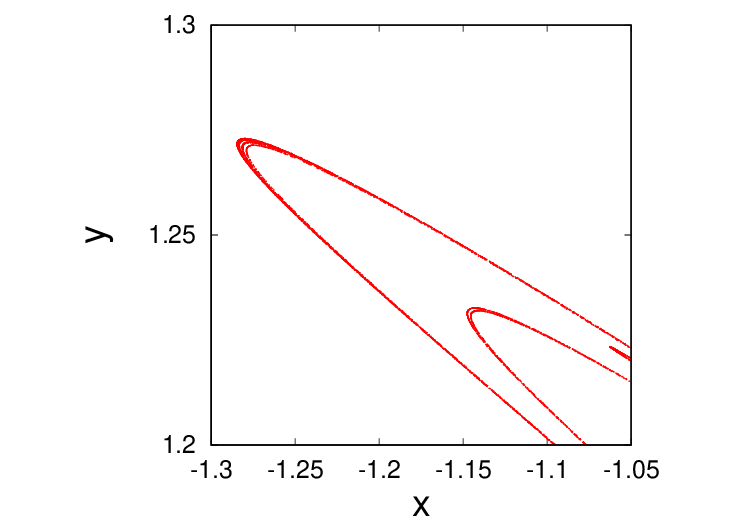
    }\hspace{-11mm}
    \includegraphics[width=0.50\columnwidth]{
        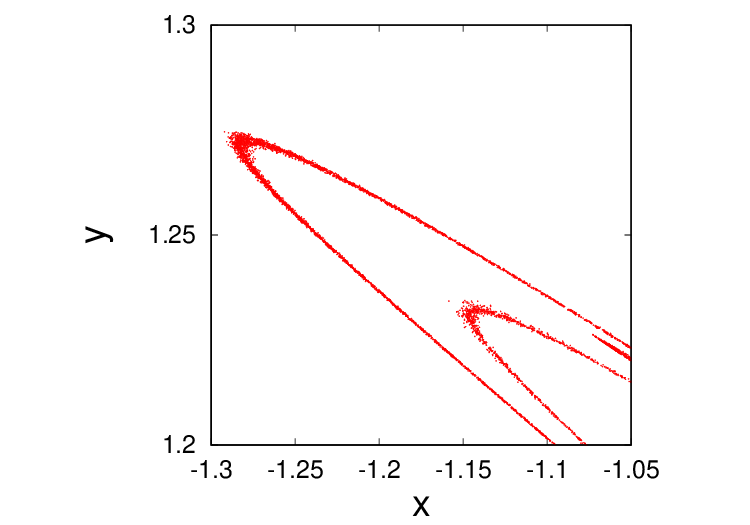
    }
    \caption{{\bf Inference of a time-series and reconstruction of the attractor.}
In the top panel, a short-term trajectory of the $x$ variable for models with spectral radius $\rho=0.1$ (left) and $0.209$ (right) in red, alongside a trajectory of the $x$ variables of the actual H\'enon map in blue.
The middle panel shows the attractor of each constructed model. The bottom panel shows an enlarged view of the upper-left part of the corresponding upper panel.
}
    \label{fig:trajectories} 
\end{figure}

\begin{figure}
    \begin{center}
        \includegraphics[width=0.95\columnwidth]{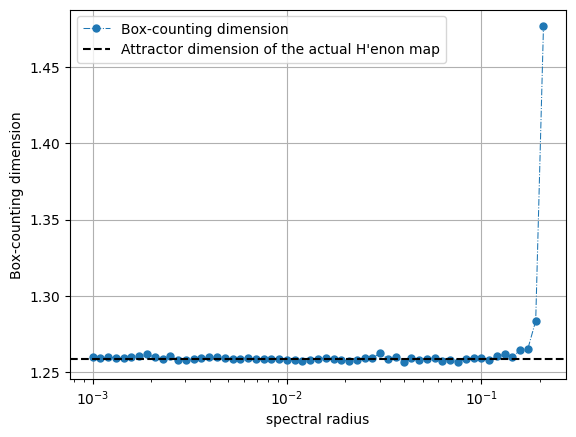}
    \end{center}
    \caption{ 
    {\bf Box counting dimension of an attractor in the reservoir space with respect to the spectral radius $\rho$.} 
    The box counting dimension of an attractor in the reservoir space is estimated from a long-term time series.
    The black dashed line represents the attractor dimension (1.258) of the actual H\'enon map.
    For the actual H\'enon map, the Lyapunov and box-counting dimensions exhibit similar values.
    The dimension of the attractor in each model is less than two.
    When $\rho$ is larger than $0.1$, the box counting dimension deviates from line (1.258), indicating that a model with a large $\rho$ generates a slightly fat attractor.
    }
    \label{fig:box-counting-dim}
\end{figure}

Figure~\ref{fig:trajectories} shows short-term trajectories and attractors for some models using various spectral radii.
While both models effectively reconstruct trajectories,  enlarged views reveal that the model with a larger spectral radius $\rho$ does not accurately reconstruct the attractor.
Figure~\ref{fig:box-counting-dim} shows the box-counting dimension \cite{falconer_1990} of data-driven models in the reservoir space for different spectral radii. 
For small spectral radii, the box-counting dimension is approximately $1.258$, 
closely matching that of the attractor of the actual H\'enon map. 
However, when the spectral radius $\rho$ exceeds $0.19$, the box-counting dimension increases more than that of the attractor of the actual H\'enon map. 
The difference in the dimensions reflects the reproducibility of the 
fractal structure in the stable direction, shown 
in the right panels of Fig.~\ref{fig:trajectories}.

\section{Geometric structure of reservoir space: Lyapunov exponents and Lyapunov vectors}
\label{sec:lyap-analysis}

    \begin{figure}
        \begin{center}
            \includegraphics[width=0.9\columnwidth, height=0.5\columnwidth]{
                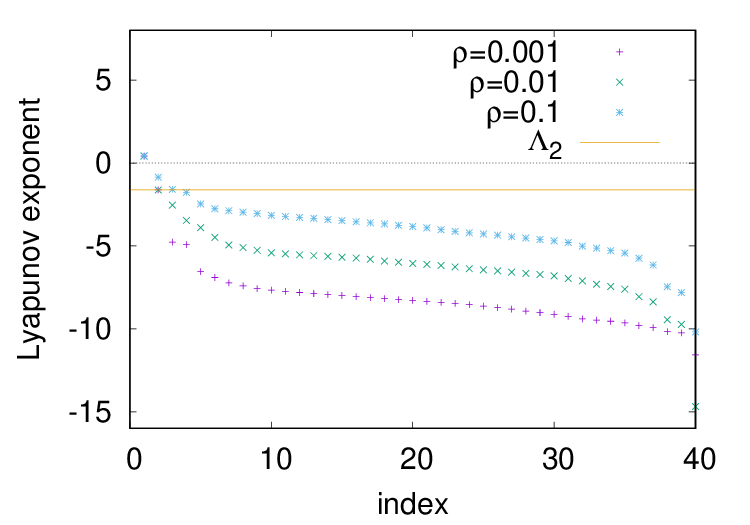
            }
        \end{center}
        \caption{{\bf Lyapunov spectrum \{$\lambda_i$\}
        of the constructed data-driven model for the cases $\rho=0.001, 0.01, 0.1$.}
        The first Lyapunov exponent $\Lambda_1$ is reconstructed for each case. For $\rho=0.001$ or $0.01$, the second Lyapunov exponent $\lambda_2(\rho)$ corresponds to the second $\Lambda_2$ of the actual H\'enon map. For $\rho=0.1$, the third Lyapunov exponent $\lambda_3(\rho)$ corresponds to $\Lambda_2$.
        All Lyapunov exponents except the first one are negative.
        }
         \label{fig:Lyapspectrum}
    \end{figure}

    \begin{figure}
        \begin{center}
            \includegraphics[width=0.9\columnwidth, height=0.5\columnwidth]{
                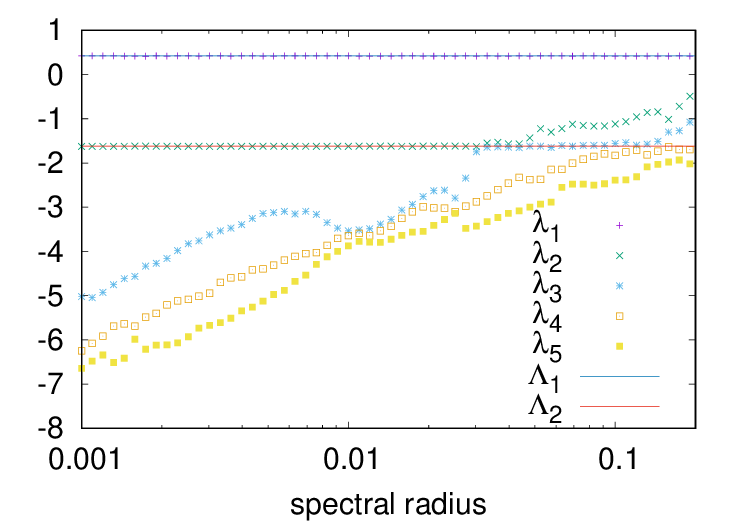
            }
        \end{center}
        \caption{{\bf Model's Lyapunov exponents $\lambda_i$ as a function of the spectral radius $\rho$ of matrix $\mb{A}$.}
        The model's first Lyapunov exponent in reservoir space $\lambda_{1}$,
        consistently reconstructs the actual first Lyapunov exponent $\Lambda_{1}$. Among
        the Lyapunov exponents in reservoir space $\lambda_{1}, \lambda_{2}, \lambda
        _{3}$, a Lyapunov exponent reconstructs the actual second
        Lyapunov exponent $\Lambda_{2}$ for models with varying spectral
        radius of matrix $\mb{A}$. \label{fig:spectrumradius} }
    \end{figure}

    \begin{figure}
        \begin{center}
            \includegraphics[width=0.95\columnwidth]{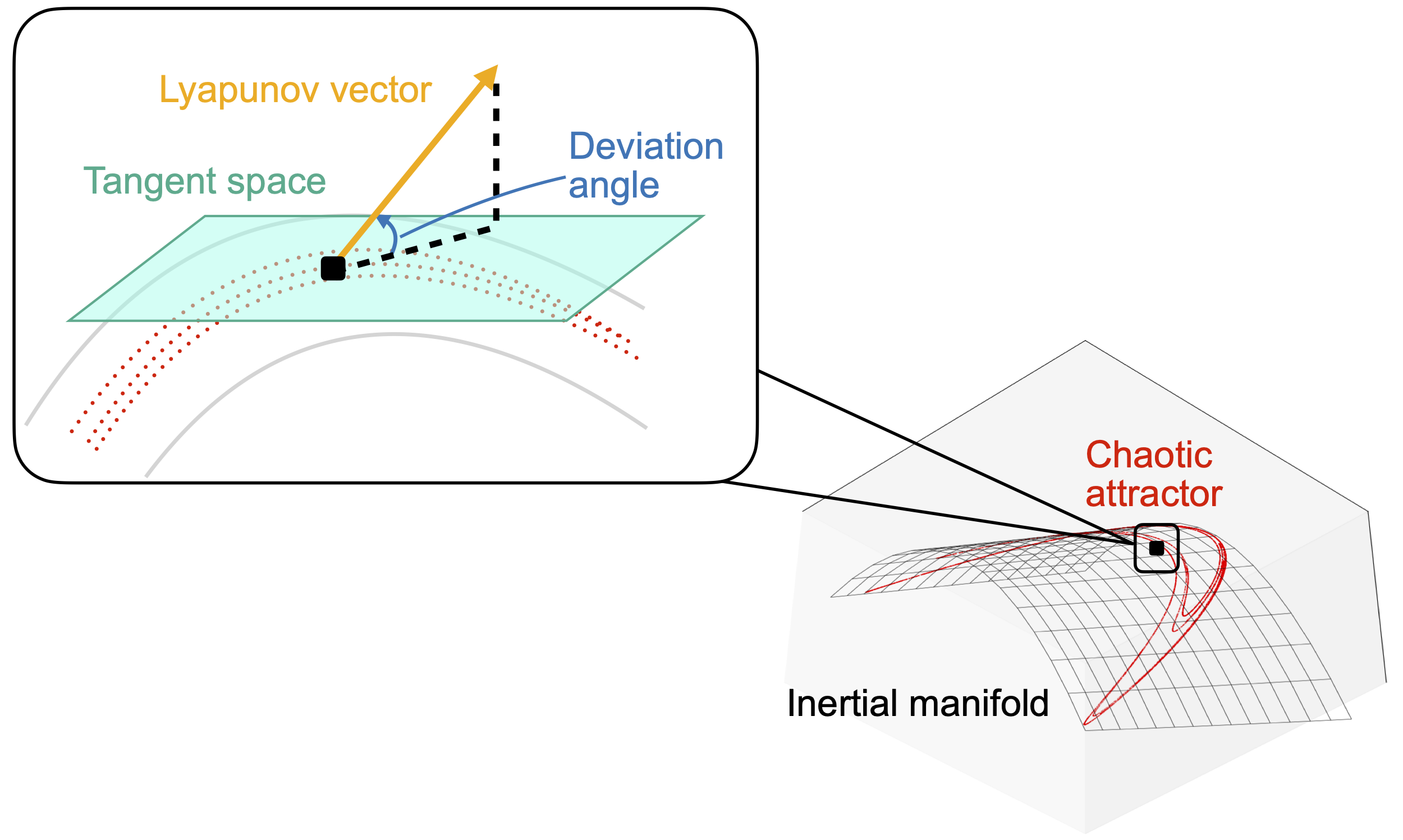}
        \end{center}
        \caption{ {\bf Schematic of the inertial manifold.} Original dynamics
        can be realized on the low-dimensional inertial manifold in the
        high-dimensional reservoir space. 
        The angle between the
        Lyapunov vector and the tangent space of the inertial manifold,
        deviation angles were calculated.
        }
        \label{fig:deviation-angle-pic}
    \end{figure}

    \begin{figure}[htbp]
        \centering
        \includegraphics[width=0.3\textwidth]{
            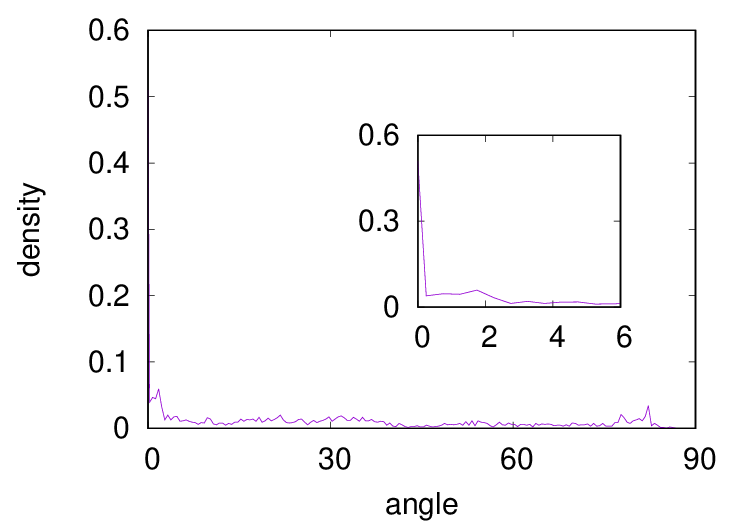
        }

        \vspace{1em} 

        \includegraphics[width=0.3\textwidth]{
            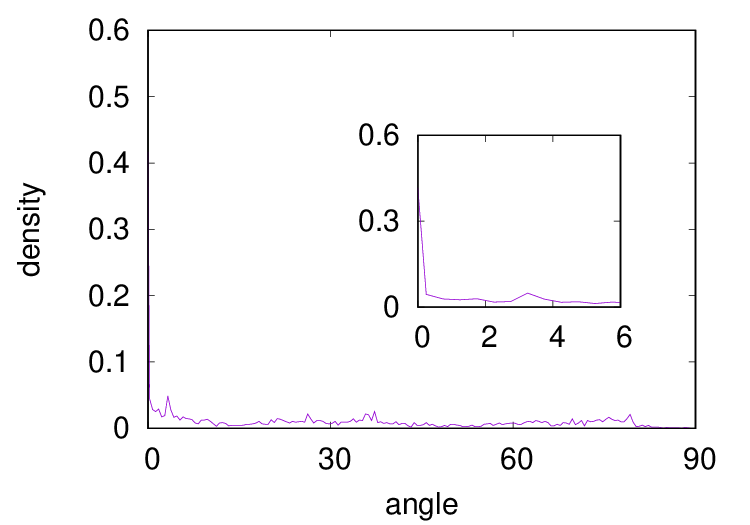
        }
        \includegraphics[width=0.3\textwidth]{
            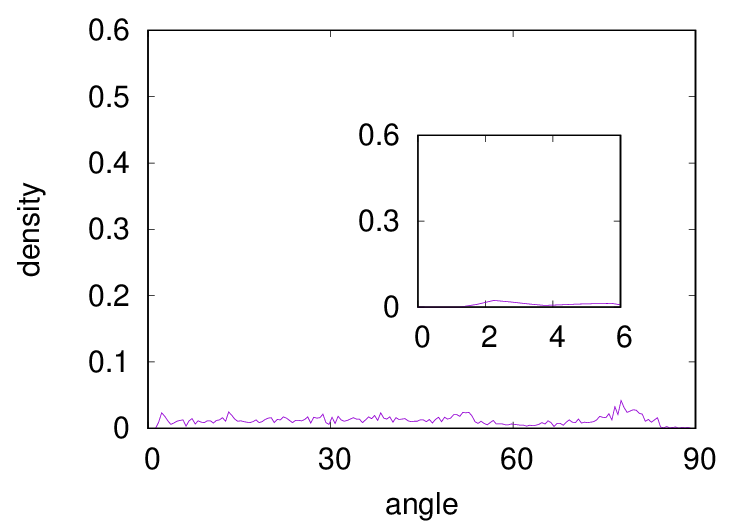
        }
        \caption{{\bf Angular distributions between the $i$-th Lyapunov vector and the tangent space of the inertial manifold along an orbit with spectral radius $\rho = 0.001$ ($i=1$ (top), $i=2$ (middle), $i=3$ (bottom)). Inset figures show enlarged views of angles near 0.}
        The Lyapunov vectors corresponding to $\lambda_1$ ($i=1$) and $\lambda_{2}$ ($i=2$) lie within the tangent space, whereas the Lyapunov vector
        corresponding to $\lambda_{3}$ ($i=3$) is transversal to the tangent space.}
        \label{fig:angular-dist}
    \end{figure}

    \begin{figure}
        \begin{center}
            \includegraphics[width=0.9\columnwidth, height=0.5\columnwidth]{
                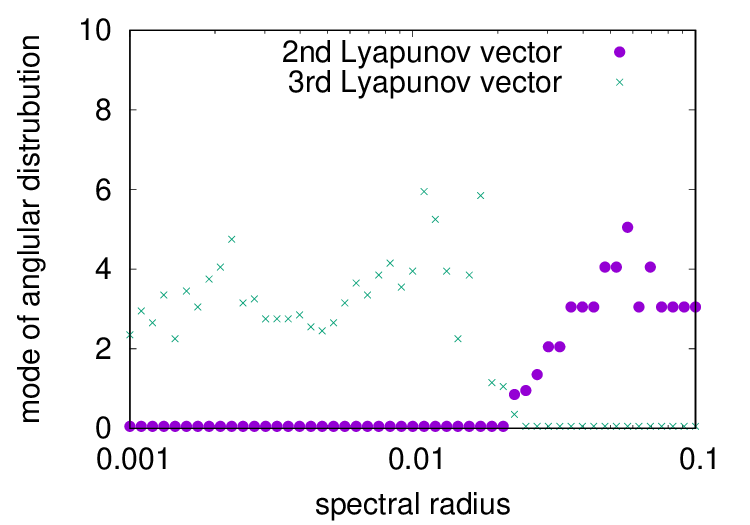
            }
        \end{center}
        \caption{{\bf Mode of angular distribution between the second and third Lyapunov vectors and a tangent space of inertial manifold with respect to the spectral radius $\rho$.} 
        The second or third Lyapunov vector lies on the inertial manifold in some spectral radii, in which each Lyapunov exponent in reservoir space corresponds to the actual second Lyapunov exponent. 
        }
    \label{fig:mode}
    \end{figure}
\subsection{Lyapunov exponents}
In this subsection, we elucidate the relationship between the spectral radius $\rho$ of the matrix ${\mb A}$ 
and the Lyapunov exponents of the model, along with the geometric structure that generates the relation. 
We calculate the Lyapunov exponents of each model using 
the QR decomposition method ~\cite{Nagashima_1969} for 
different spectral radii $\rho$ in the reservoir space and denote the $i$-th exponents as $\lambda_{i}(\rho)$.
The Lyapunov exponents of the actual H\'enon map, termed the actual Lyapunov exponents, are described as $\Lambda_i~(i=1,2)$, which are $(\Lambda_1, \Lambda_2) \approx (0.419, -1.623)$.
Figure~\ref{fig:Lyapspectrum} shows the Lyapunov spectra for data-driven models with $\rho=0.001, 0.01, 0.1$.
The $i$-th Lyapunov exponent $(5\leq i \leq 35)$ in the case with $\rho=0.1$ (respectively, $\rho=0.01$) is approximately $\log_e 10 (\approx 2.3)$ times larger than the $i$-th Lyapunov exponent $\lambda_i$ $(5\leq i \leq 35)$ in the case with $\rho=0.01$ (respectively, $\rho=0.001$). 
 These Lyapunov exponents are primarily determined by the trace of the matrix ${\bf A}$ of the model, indicating that the spectral radius $\rho$ dominates the exponents.
The first Lyapunov exponent $\lambda_1(\rho)$ for all $\rho$ corresponds to the one $\Lambda_1$ for the actual model.
The second Lyapunov exponent $\lambda_2(\rho)$ in the case with $\rho=0.001, 0.01$ corresponds to the second one $\Lambda_2$ of the actual model, while for $\rho=0.1$, the third Lyapunov exponent $\lambda_3$ corresponds to the second $\Lambda_2$ in the actual model.
Figure~\ref{fig:spectrumradius} shows the first to the fifth Lyapunov exponents of constructed models across various spectral radii $\rho$.
For all $\rho$, the first Lyapunov exponent $\lambda_{1}(\rho)$ closely matches the actual positive Lyapunov exponent.
The remaining Lyapunov exponents of the models are negative.
For any $\rho$, one Lyapunov exponent approximates the negative Lyapunov exponent of the original dynamics. 
For models with small $\rho$, the second Lyapunov exponent $\lambda_{2}(\rho)$ is close to the actual negative exponent $\Lambda_2$, indicating strong contraction in transversal directions. 
For large $\rho$, the second Lyapunov exponent $\lambda_{2}(\rho)$ is larger than the negative Lyapunov exponent of the original dynamics.
All Lyapunov exponents, except those corresponding to the first and second Lyapunov exponents of the actual H\'enon map,
are negative, with their
corresponding vectors expected to be directed in the transversal direction to the attractor.
The original dynamics are anticipated to be reconstructed on a low-dimensional inertial manifold.
This structure is supported by several studies~\cite{hart24}.
We numerically validate this fact in the next subsection.

\subsection{Lyapunov vectors}
We investigate the structure and existence of the low-dimensional inertial manifold using the covariant Lyapunov vector~\cite{ginelli_2007,saiki_2010}, which indicates the directions of perturbation growth corresponding to each Lyapunov exponent.
The two-dimensional tangent space of the manifold can be numerically estimated from long-term time series by selecting two trajectory points near a targeted point and applying the Gram-Schmidt process. 
Details of the algorithm for constructing the tangent space are provided in Section~\ref{subsec:algorithm}. 
We calculate the deviation angle between the Lyapunov vector and the tangent space of the manifold~(See Fig.~\ref{fig:deviation-angle-pic}). 
A non-zero angle indicates that the Lyapunov vector is transverse to the manifold. 
Figure~\ref{fig:angular-dist} shows the density distributions of the deviation angles for the model with $\rho=0.001$, where the first and second Lyapunov exponents $\lambda_1, \lambda_2$, closely match the actual Lyapunov exponents.
The distributions for the first and second Lyapunov exponents, which are similar to the actual Lyapunov exponents, peak at zero, whereas those for other Lyapunov exponents do not. 
These results indicate that, for $\rho=0.001$, the Lyapunov exponents, which are not similar to the actual Lyapunov exponents, correspond to the transversal direction, and the dynamics are embedded in a two-dimensional manifold. 
We validate this structure for various $\rho$ values.
As an indicator of how transversely the Lyapunov vectors are oriented, we use the mode of the deviation angle distribution.
When the indicator is far from $0$, the corresponding Lyapunov vector is transversal.
%Figure~\ref{fig:mode} shows the indicators for the third Lyapunov exponents across various $\rho$. 
Figure~\ref{fig:mode} shows the indicators for the second and third Lyapunov exponents across various $\rho$. 
For small $\rho$, the third Lyapunov exponent takes a different value from the actual Lyapunov exponents, and the corresponding Lyapunov vector is transversal.
When the third Lyapunov exponent matches the actual Lyapunov exponents, the indicator is $0$, indicating that the vector lies within the tangent space of the two-dimensional inertial manifold.
Thus, the Lyapunov vector with the exponent similar to the actual exponent lies in the tangent space of the manifold, while others are transversal, confirming the embedding structure.

\section{Lyapunov exponents on the inertial manifold}
In this section, we calculate the Lyapunov exponents in the inertial manifold.
The original dynamics are reconstructed in the inertial manifold.
When determining the Lyapunov exponents in the manifold, the actual Lyapunov exponents are also calculated, 
providing further evidence of the embedding structure.
We denote the Lyapunov exponents constrained on the inertial manifold of the model as $\tilde{\lambda}_{i}~(i=1,2$ for the actual H\'enon map).

\subsection{Algorithm}
\label{subsec:algorithm}
    \begin{figure}
        \begin{center}
            \includegraphics[width=0.95\columnwidth]{
                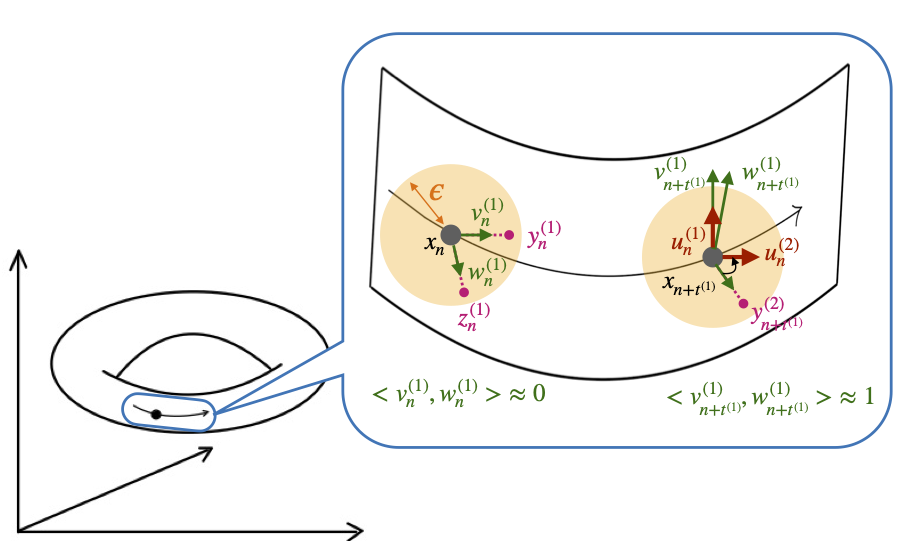
            }
        \end{center}
        \caption{ {\bf Schematic of identifying the Gram--Schmidt Lyapunov vectors in the tangent space of the two-dimensional inertial manifold.}
        Using the Lyapunov vectors, we estimated Lyapunov exponents restricted to
        the tangent space spanned using the numerically computed trajectory of the model.
        }
         \label{fig:algorithm-sketch}
    \end{figure}

We introduce an algorithm to calculate the Lyapunov exponents in the inertial manifold by estimating the Gram--Schmidt Lyapunov vectors and calculating their corresponding Lyapunov exponents.
The dimension of the inertial manifold $D$ can be estimated from the time series using a box-counting dimension. 
The schematic for $D=2$ is shown in Fig.~\ref{fig:algorithm-sketch}.

We focus on a sample point $x_{n}$ in a long trajectory to estimate
the Lyapunov vectors and denote its $\epsilon$-neighborhood of $x_{n}$ as $U(x_{n})$.
To estimate the first Lyapunov vector, we select two points $y_{n}^{(1)}, z_{n}^{(1)}\in U(x_{n})$ in the long trajectory satisfying $\langle v_{n}^{(1)},w_{n}^{(1)}\rangle \ < \  \delta_1 \approx 0$, where $v_{n}^{(1)}:=(y_{n}^{(1)}-x_{n})/|y_{n}^{(1)}-x_{n}|, w_{n}^{(1)}:=(z_{n}^{(1)}-x_{n})/|z_{n}^{(1)}-x_{n}|$, $\langle \cdot,\cdot \rangle$ is the standard inner product, and $| \cdot|$ is the standard norm. 
For a value $\delta_2$ close to 1,
we follow the trajectories until $\langle v_{n+t^{(1)}}^{(1)},w_{n+t^{(1)}}^{(1)}\rangle \ > \  \delta_2$. 
We define $v_{n+t^{(1)}}^{(1)}$ as the first Lyapunov vector in the low-dimensional inertial manifold, denoted as 
$u^{(1)}_n$.

To estimate the $i$-th Gram--Schmidt Lyapunov vector in the low-dimensional inertial manifold~($i=2,\ldots, D-1$), the following algorithm was applied. 
We designate the targeted point $x_{n+t^{(i-1)}}$ as $x_{n}$.
We select two points $y_{n}^{(i)}, z_n^{(i)} \in U(x_{n})$ satisfying $\langle v_{n}^{(i)},w_{n}^{(i)}\rangle < \delta_1$, where the directions of $\{u^{(j)}\}_{j=1, \ldots i-1}$ are removed from $y_{n}^{(i)}$ 
and define the vector as $v_{n}^{(i)}$ after normalizing, and apply the same procedure for $z_{n}^{(i)}$ instead of $y_{n}^{(i)}$ and define the applied vector as $w_{n}^{(i)}$. 
The trajectories are also followed until $\langle v_{n+t^{(i)}}^{(i)},w_{n+t^{(i)}}^{(i)}\rangle \ > \ \delta_2$. We define $v_{n+t^{(i)}}^{(i)}$ as the $i$-th Lyapunov vector in the low-dimensional inertial manifold, which we denote as $u^{(i)}_n$.

To estimate the $D$-th Gram--Schmidt Lyapunov vector, we select a point $y_{n}^{(D)}\in U(x_{n})$. The $D$-th Lyapunov vector $u^{(D)}$ is defined as $y_{n}^{(D)}$ orthogonalized with respect to $\{u^{j}_n\}_{j=1, \ldots D-1}$. 

\subsection{Results}

% \begin{table}[tb]
%     \footnotesize
%     \centering
%     \begin{tabular}{|c|r|r|r|}
%         \hline
%         & \multicolumn{1}{c|}{$\epsilon$} & \multicolumn{1}{c|}{$\delta_{1}$} & \multicolumn{1}{c|}{$\delta_{2}$}\\
%         \hline
%         H\'enon map & $10^{-3}$  & $0.3$  & $0.999$ \\
%         \hline
%     \end{tabular}
%     \caption{
%     {\bf Parameters for estimating the Lyapunov exponents in the algorithm.}
%     }
%     \label{tab:parameter_estimation}
% \end{table}

\begin{table}[tb]
    % \footnotesize
    \centering
    \begin{tabular}{|c|c|c|}
        \hline
        $\epsilon$ & $\delta_{1}$ & $\delta_{2}$\\
        \hline
         $10^{-3}$  & $0.3$  & $0.999$ \\
        \hline
    \end{tabular}
    \caption{
    {\bf Parameters for estimating the Lyapunov exponents in the algorithm.}
    }
    \label{tab:parameter_estimation}
\end{table}

\begin{figure}[t]
    \begin{center}
        \includegraphics[width=0.9\columnwidth, height=0.5\columnwidth]{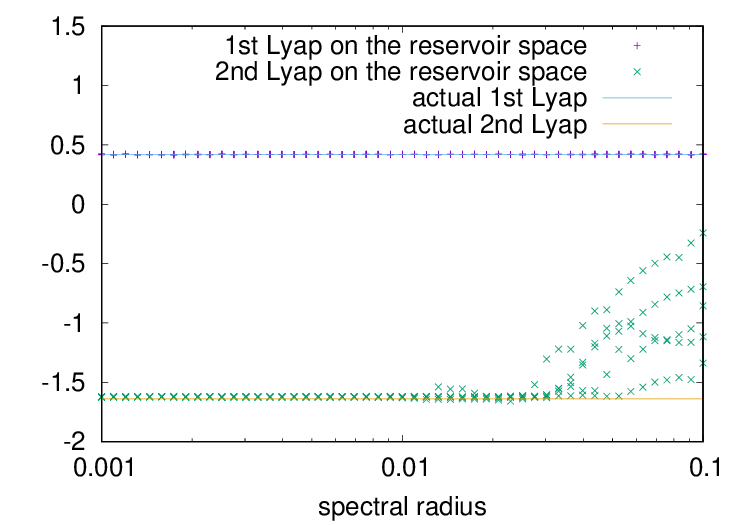}
        \includegraphics[width=0.9\columnwidth, height=0.5\columnwidth]{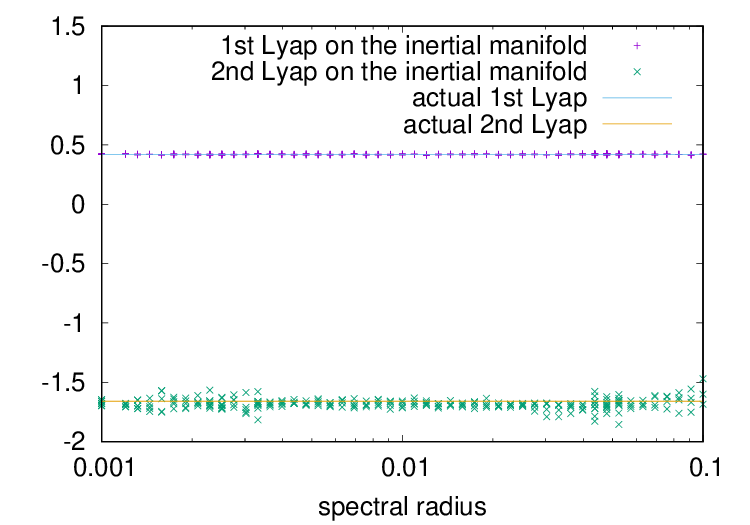}
    \end{center}
    \caption{
    {\bf The first and second Lyapunov exponents calculated in the reservoir space and in the manifold for a model concerning the various spectral radii $\rho$.}
    %, respectively.}
    The top panel indicates the first and second Lyapunov exponents in reservoir space($\lambda_{1}$($+$) and $\lambda_{2}$ ($\times$)).
    The first Lyapunov exponent $\lambda_{1}$ reconstructs the actual one $\lambda_{1}$ for each of the cases with high accuracy (error is less than 1\%). 
    The second Lyapunov exponent $\lambda_{2}$ reconstructs the actual one $\Lambda_{2}$ only when the spectral radius is small ($\rho \leq 0.01$).
    The bottom panel represents the first and second Lyapunov exponents in the inertial manifold ($\tilde{\lambda}_{1}$($+$) and $\tilde{\lambda}_{2}$($\times$)).
    The second Lyapunov exponent $\tilde{\lambda}_{2}$ is estimated to be approximately the actual value $\Lambda_{2}$, even when the spectral radius is large ($\rho > 0.01$).
    In both panels, for each $\rho$, five models are used, including the model used in Section~\ref{sec:lyap-analysis}.
    } 
    \label{fig:2nd-lyap-sub}
\end{figure}

We calculate the Lyapunov exponents constrained on the inertial manifold $\tilde{\lambda}_{i}$ for models with various spectral radii using the method explained in Section~\ref{subsec:algorithm}.
Fig.~\ref{fig:2nd-lyap-sub} shows the Lyapunov exponents for five models, each constructed with different random matrices $\mb{{A^\prime}}$ and $\mb{W}_{\text{in}}$.
The first Lyapunov exponents $\tilde{\lambda}_1$ closely match the actual positive Lyapunov exponent $\Lambda_1$ for any spectral radius, consistent with unrestricted Lyapunov exponents $\lambda_1$ in the reservoir space.
The second Lyapunov exponents $\tilde{\lambda}_2$ are similar to the actual negative Lyapunov exponents $\Lambda_2$, even when the second Lyapunov exponents in the reservoir space $\lambda_2$ are larger than the actual negative value.
Since the Lyapunov vectors corresponding to Lyapunov exponents are different from the actual Lyapunov exponents are transversal, they do not affect the expansion rate on the manifold, allowing the Lyapunov exponents on the manifold to robustly take the actual value.
These results confirm the existence of the inertial manifold, where the original dynamics are reconstructed.

\section{Concluding remarks}
\subsection{Summary of results}
For any spectral radius $\rho$, the Lyapunov exponent for the reservoir computing model includes the actual first and second Lyapunov exponents, with all other Lyapunov exponents of the reservoir computing model being negative. 
We investigated the change in the negative Lyapunov exponent (the second Lyapunov exponent for the actual H\`enon map) in models by varying the spectral radius $\rho$.
As shown in Fig.~\ref{fig:spectrumradius}, when the spectral radius $\rho$ is sufficiently small ($\rho\approx 0.001$), the second Lyapunov exponent in the reservoir space matches that of the original dynamical system. 
As the spectral radius $\rho$ increases, the model's second Lyapunov exponent becomes larger and deviates from the original second Lyapunov exponent, but the third Lyapunov exponent aligns with the original second Lyapunov exponent.
This indicates that structural properties of the original system are reconstructed robustly in the low dimensional inertial manifold embedded in high dimensional neural network.

To validate the structure, we calculated the deviation angle, defined as the angle between the tangent space of the inertial manifold and each Lyapunov vector. 
Figure~\ref{fig:angular-dist} shows the distribution of deviation angles between the third Lyapunov vector and the tangent space of the inertial manifold, estimated from the time series data at each trajectory point.
For a small spectral radius, the third Lyapunov exponent of the reservoir space is sufficiently small and oriented transversely,
supporting the existence of an inertial manifold where the actual dynamics are reconstructed in a reservoir space.
To further confirm this structure, we calculated the Lyapunov exponents in the manifold, as shown in Fig.~\ref{fig:2nd-lyap-sub}, which can be obtained even when Lyapunov exponents of the original dynamics are excluded from those of the constructed model.
The second Lyapunov exponents computed in the manifold remain similar to those of the original dynamics, even when the spectral radius $\rho$ is large.

These findings suggest that the spectral radius $\rho$ has a significant impact on the model's dynamical properties, particularly the Lyapunov spectrum, and that reservoir models can accurately reconstruct the underlying characteristics of the original system. 

\subsection{Discussion} 
In summary, when reservoir computing accurately reproduces trajectories, the original dynamics are reconstructed on a low-dimensional manifold within the reservoir space.
Similar geometric structures are discussed in some theoretical studies on dynamical systems~\cite{tempkin2007, ott2003}.

When an original attractor is embedded in a high-dimensional space, its reconstructed trajectory resides on a low-dimensional manifold, where the Lyapunov exponents of the original attractor are reproduced~\cite{ott2003}.
This structure is realized in reservoir computing, indicating that the original attractor is embedded on a low-dimensional manifold.
However, when an original attractor is embedded, transversal directions to the tangent space of the manifold may be unstable, indicating that the reconstructed invariant set is not an attractor in high-dimensional space~\cite{tempkin2007}.

Even when the original attractor is embedded in a low-dimensional manifold, the model's numerical trajectories are less accurate if the transversal stability is weak.
The lack of stability in the transversal directions causes deviations from the original dynamics. 
If the transversal directions are unstable, we cannot generate a proper long-term numerical trajectory.
For effective trajectory reproduction in applications, both embedding the original attractor on the manifold and stabilizing transversal directions are crucial.
The generalized synchronization facilitates this stabilization.
In reservoir computing, the embedding of original dynamics is robust across various spectral radii $\rho$, which primarily governs the stability of transversal directions.
To accurately reproduce original dynamics as an autonomous system using reservoir computing, it is important to stabilize the transversal directions by adjusting the spectral radius. 
For two-dimensional dynamics, a constructed model achieves high accuracy when the transversal contraction is sufficiently strong, such that $\Lambda_2 > \lambda_n (n=3, \ldots, N)$.

Even when transversal directions are unstable, preventing the generation of accurate long-term trajectories through forward iteration of the model, the Stagger-and-Step method can still produce reliable trajectories~\cite{sweet_2001c}.
Similarly, a data-driven model employing an alternative modeling method to reservoir computing can generate accurate trajectories when the original attractor is embedded, despite exhibiting transverse instability~\cite {tsutsumi23, tsutsumi25}.

\section*{Acknowledgements}
KN was supported by the JSPS KAKENHI Grant 	No. 22K17965. 
%Project of President Discretionary Budget of TUMST. 
YS was supported by the JSPS KAKENHI Grant No. 25H01469 and 24K00537.
%No.19KK0067 and 21K18584, and JSPS
%Bilateral Open Partnership Joint Research Projects JPJSBP 120229913. 
Part of the computation was supported by JHPCN (h250021).
%(jh210027,jh220007), HPCI (hp210072),
and the Collaborative Research Program for Young $\cdot$ Women Scientists of ACCMS and IIMC, Kyoto University.

\section*{Data Availability Statement}
The data and code that support the findings of this study are available from the corresponding author upon reasonable request.

\section*{Conflict of interest}
The authors have no conflicts to disclose.

% \clearpage
%\references
\section*{References}
\bibliographystyle{Science}
\bibliography{reservoir_bibliography,qr_bibliography,heterochaos_bibliography,heterochaos-reservoir}
%\bibliography{output} 
\end{document}